\documentclass[10pt]{article}

\usepackage{hyperref}
\usepackage{amscd}
\usepackage{amsmath}
\usepackage{amsfonts}
\usepackage{amssymb}
\usepackage[english]{babel}

\begin{document}

\title{A Note on Hamming distance of constacyclic codes of length $p^s$ over $\mathbb F_{p^m} + u\mathbb F_{p^m}\footnote{
 E-Mail addresses: hwliu@mail.ccnu.edu.cn (H. Liu), m.youcef@mails.ccnu.edu.cn (M. Youcef).}$ }

\author{Hongwei Liu$^1$,~Maouche Youcef$^{1,2}$}

\date{\small
${}^1$School of Mathematics and Statistics, Central China Normal University,Wuhan, Hubei, 430079, China\\
${}^2$Department of Mathematics, University of Sciences and Technology HOUARI BOUMEDIENE, Alger, Algeria\\
}
\maketitle

\leftskip 0.8in
\rightskip 0.8in
\noindent
{\bf Abstract.}  For any prime $p$, $\lambda$-constacyclic codes of length $p^s$ over ${\cal R}=\mathbb{F}_{p^m} + u\mathbb{F}_{p^m}$ are precisely the ideals of the local ring ${\cal R}_{\lambda}=\frac{{\cal R}[x]}{\left\langle x^{p^s}-\lambda \right\rangle}$, where $u^2=0$. In this paper, we first investigate the Hamming distances of cyclic codes of length $p^s$ over ${\cal R}$. The minimum Hamming distances of all cyclic codes of length $p^s$ over ${\cal R}$ are determined. Moreover, an isometry between cyclic and $\alpha$-constacyclic codes of length $p^s$ over ${\cal R}$ is established, where $\alpha$ is nonzero elements of $\mathbb{F}_{p^m}$, which carries over the results regarding cyclic codes corresponding to $\alpha$-constacyclic codes of length $p^s$ over ${\cal R}$.



\leftskip 0.0in
\rightskip 0.0in

\vskip 30pt

\section{Introduction}
Constacyclic codes over finite fields are generalizations of cyclic and negacyclic codes, and it's form a well-known family of linear codes,
which form an important class of linear codes containing many optimal codes. However, most of the research is concentrated on the situation
when the code length $n$ is relatively prime to the characteristic of the field $F$. The case where the code length $n$ is not relatively prime to
 the characteristic of $F$ yields the so-called repeated-root codes. which were first studied by Berman \cite{berman1967}, and then by several authors,
 such as Massey et al. \cite{massey1973}, Falkner et al. \cite{falkner1979}, and Roth and Seroussi \cite{roth1986}. Repeated-root codes were investigated in the
 most generality in the 1990s by Castagnoli et al. \cite{castagnoli1991}, and van Lint \cite{van1991}, where they showed that repeated-root cyclic codes have a
 concatenated construction, and are asymptotically bad. Nevertheless, such codes are optimal in a few cases,
 that motivates researchers to further study this class of codes (see, for example, \cite{tang1997}, \cite{nedeloaia2003}, \cite{zimmermann}).

\vskip 6pt
After the realization \cite{calderbank}, \cite{hammons1994}, \cite{nechaev1991}, that many seemingly nonlinear binary codes are actually closely related to linear codes over the ring $\mathbb{Z}_4$, codes over $\mathbb{Z}_4$ in particular, and codes over finite rings in general, have received a great deal of attention. The ring $\mathbb{F}_{2} + u\mathbb{F}_{2}$, where $u^2=0$, is interesting because it additively similarly to $\mathbb{F}_{4}$, and multiplicatively similarly to $\mathbb{Z}_{4}$. $(1-u)$-constacyclic codes over $\mathbb{F}_{2} + u\mathbb{F}_{2}$ of odd length were first introduced by Qian et al in\cite{qian2006}, where it proved that the Gray image of a linear $(1-u)$-constacyclic code over $\mathbb{F}_{2} + u\mathbb{F}_{2}$ of odd length is a binary distance invariant linear cyclic code. Later, in \cite{amarra2008}, Amarra and Nemenzo extended the results to codes over $\mathbb{F}_{p^k} + u\mathbb{F}_{p^k}$,  and studied $(1-u)$-constacyclic codes over $\mathbb{F}_{p^k} + u\mathbb{F}_{p^k}$ of length $n$ under the condition $\gcd(n,p)=1$.

\vskip 6pt
In 2009, Dinh \cite{dinh2009} studied the algebraic structure of repeated-root $(\alpha+u\beta)$-constacyclic codes over $\mathbb{F}_{2^m} + u\mathbb{F}_{2^m}$ of length $2^s$, and determined the Hamming distances of of all such codes under the condition $\beta \neq 0$. In recent years, Dinh extended the results of \cite{dinh2009} to codes over ${\cal R}=\mathbb{F}_{p^m} + u\mathbb{F}_{p^m}$, and obtained the number of codewords in each of those cyclic codes (see \cite{dinh2010}) under the same condition $\beta \not =0$. If $\beta =0$ Dinh showed that, unlike the chain ring ${\cal R}_{\alpha+u\beta}$, the ring ${\cal R}_\alpha$ is a local ring with maximal ideal $\langle 2, x-\alpha_0\rangle$ but not a chain ring, hence, the structure of the rings ${\cal R}_\alpha$ are more complicated, it follows that the Hamming distance is more difficult to compute than that of the chain rings ${\cal R}_{\alpha+u\beta}$ case.

\vskip 6pt
In this paper, we continue to study Hamming distance of $\alpha$-repeated-root constacyclic codes over the rings $\mathbb{F}_{p^m}+u\mathbb{F}_{p^m}$ under the rest condition ($\beta = 0$), where the codes are exactly the ideals of the local ring ${\cal R}_\alpha$, which is a local ring with maximal ideal $\langle 2, x-\alpha_0\rangle$, but not a chain ring. After giving some preliminaries in Section 2, the rest of the paper is organized as follows. In Section 3, we proceed by first giving the structure and Hamming distances of all constacyclic codes of length $p^s$ over the finite field $\mathbb{F}_{p^m}$, then we give the structure of $\alpha$-constacyclic codes of length $p^s$ over $\mathbb{F}_{p^m}+u\mathbb{F}_{p^m}$ obtained in \cite{dinh2010}. Section 4 addresses the cyclic codes of length $p^s$ over ${\cal R}$, we establish the Hamming distances of cyclic codes over ${\cal R}$. In Section 5, we construct a one-to-one rings isomorphism between ${\cal R}_1$ and ${\cal R}_\alpha$ which allows us to apply our results about cyclic codes to $\alpha$-constacyclic codes over ${\cal R}$.

\vskip 30pt
\section{Preliminaries}

Let $R$ be a finite commutative ring. An ideal $I$ of $R$ is called principal if it is generated by one element. $R$ is called a principal ideal ring if all the ideals of $R$ are principal. The ring $R$ is called a local ring if $R$ has a unique maximal ideal. Furthermore, $R$ is said to be a
chain ring if it is a local principal ideal ring. The following result is well known (cf. [\cite{dinh2004},Proposition 2.1]).

\vskip 6pt
\noindent {\bf Proposition 2.1.} {\it Let $R$ be a finite commutative ring, then the following conditions are equivalent:

{$(i)$} $R$ is a local ring and the maximal ideal $M$ of $R$ is principal, i.e., $M=\langle \gamma \rangle$ for some $\gamma\in R$,

{$(ii)$} $R$ is a local principal ideal ring,

{$(iii)$} $R$ is a chain ring whose ideals are $\langle \gamma^i \rangle$, $0 \leq i \leq e$, where $e$ is the nilpotency index of $\gamma$.

\vskip 4pt
Moreover, if $R$ is a finite commutative chain ring with maximal ideal $\langle\gamma\rangle$, and the nilpotency index of $\langle\gamma\rangle$ is $e$,  then the cardinality of the ideal $\langle \gamma^i \rangle$ is $|R/\langle \gamma \rangle|^{e-i}$, where $i=0,1,\cdots, e-1$.}

\vskip 6pt
Let $R$ be a finite ring, a code $C$ of length $n$ over $R$ is a nonempty subset of $R^n$, and the ring $R$ is referred to as the alphabet of the code. If this subset is, in addition, a $R$-submodule of $R^n$, then $C$ is called {\it linear}. For a unit $\lambda$ of $R$, the $\lambda$-constacyclic ($\lambda$-twisted) shift $\tau_{\lambda}$ on $R^n$ is the shift
$$\tau_{\lambda}((x_0,x_1,\dots,x_{n-1}))=(\lambda x_{n-1},x_0,x_1,\dots,x_{n-2}),$$
and a code $C$ is said to be $\lambda$-constacyclic if $\tau_{\lambda}(C)=C$, i.e., if $C$ is closed under the the $\lambda$-constacyclic shift $\tau_{\lambda}$. In case $\lambda=1$, those $\lambda$-constacyclic codes are called cyclic codes, and when $\lambda=-1$, such $\lambda$-constacyclic codes are called negacyclic codes.

\vskip 6pt
Each codeword $c=(c_0,c_1,\dots,c_{n-1})\in C$ is customarily identified with its polynomial representation $c(x)=c_0+c_1x+\dots+c_{n-1}x^{n-1}$, and the code $C$ is in turn identified with the set of all polynomial representations of its codewords. Then in the ring $\frac{R[x]}{\langle x^n-\lambda \rangle}$, $xc(x)$ corresponds to a $\lambda$-constacyclic shift of $c(x)$. From that, the following fact is well known (cf. \cite{huffman2010, macWilliams1977}) and straightforward:

\vskip 6pt
\noindent
{\bf Proposition 2.2.} {\it A linear code $C$ of length $n$ is $\lambda$-constacyclic over $R$ if and only if $C$ is an ideal of $\frac{R[x]}{\langle x^n-\lambda \rangle}$.}

\vskip 6pt
Given $n$-tuples $x = (x_0, x_1, \dots, x_{n-1}), y = (y_0, y_1, \dots, y_{n-1}) \in R^n$, their inner product or dot product is defined as usual
$$x \cdot y = x_0y_0 + x_1y_1 + \dots +x_{n-1}y_{n-1},$$
evaluated in $R$. Two $n$-tuples $x, y$ are called {\it orthogonal} if $x \cdot y = 0$. For a linear code $C$ over $R$, its {\it dual code} $C^{\perp}$ is the set of $n$-tuples over $R$ that are orthogonal to all codewords of $C$, i.e.,
$$C^{\perp} = \{x \ | \ x \cdot y = 0, \forall y \in C \}.$$
A code $C$ is called {\it self-orthogonal} if $C \subseteq C^{\perp}$, and it is called {\it self-dual} if $C = C^{\perp}$. The following result is well known (cf. \cite{ ding2005, huffman2010, macWilliams1977, pless1998}).

\vskip 6pt
\noindent {\bf Proposition 2.3.}{\it Let $p$ be a prime and $R$ be a finite chain ring of size $p^{\alpha}$. The number of codewords in any linear code $C$ of length $n$ over $R$ is $p^k$, for some integer $k \in \{0, 1, \dots, \alpha n\}$. Moreover, the dual code $C^{\perp}$ has $p^l$ codewords, where $k+l={\alpha}n$, i.e., $|C| \cdot |C^{\perp}| = |R|^n$.}

\vskip 6pt
In general, we have the following implication of the dual of a $\lambda$-constacyclic code.

\vskip 6pt
\noindent {\bf Proposition 2.4.} {\it The dual of a $\lambda$-constacyclic code is a $\lambda^{-1}$-constacyclic code.}

\vskip 6pt
For a word $x=(x_0, x_1, . . . , x_{n-1}) \in R^n $, the Hamming weight of $x$, denoted by $wt(x)$, is the number of nonzero components of $x$. The Hamming distance $d(x, y)$ of two words $x$ and $y$ equals the number of components in which they differ, which is the Hamming weight $wt(x-y)$ of $x-y$. For a nonzero linear code $C$, the minimum Hamming weight $wt(C)$ and the minimum Hamming distance $d(C)$ are the same and defined as the smallest Hamming weight of nonzero codewords of $C$:$$d(C) = min\left\{ wt(x)\, | \,0\neq x \in C\right\}.$$
\vskip 6pt
The zero code is conventionally said to have Hamming distance 0.

In this paper, we consider all constacyclic codes of length $p^s$ with alphabet $\mathbb{F}_{p^m} + u\mathbb{F}_{p^m}$, consists of all $p^m$-ary polynomials of degree $0$ and $1$ in an indeterminate $u$, and it is closed under $p^m$-ary polynomial addition and multiplication modulo $u^2$. Thus, ${\cal R}=\frac{\mathbb{F}_{p^m}[u]}{\left\langle u^2 \right\rangle}=\left\{a+ub | a,b \in \mathbb{F}_{p^m} \right\}$ is a local ring with maximal ideal $u\mathbb{F}_{p^m}$. Therefore, by Proposition 2.1, it is a chain ring. The ring ${\cal R}$  has precisely $p^m\left(p^m-1\right)$ units, which are of the forms $\alpha+u\beta$ and $\gamma$ , where $\alpha$, $\beta$, and $\gamma$ are nonzero elements of the field $\mathbb{F}_{p^m}$.

\vskip 30pt
\section{Cyclic codes of length $p^{s}$ over $\mathbb{F}_{p^{m}}+u\mathbb{F}_{p^{m}}$}

In \cite{dinh2008}, the algebraic structure and Hamming distances of cyclic codes of length $p^s$ over $\mathbb{F}_{p^m}$ were established and given by the following theorem.

\vskip 6pt
\noindent
{\bf Theorem 3.1.}\label{Theorem-1}\cite{dinh2008}
Let $C$ be a cyclic code of length $p^{s}$ over $\mathbb{F}%
_{p^{m}}$. Then $C=\left\langle (x-1)^{i}\right\rangle
\subseteq \frac{\mathbb{F}_{p^{m}}[x]}{\left\langle x^{p^{s}}-1 \right\rangle },$ for $i\in \{0,1,...,p^{s}\}$, and its Hamming distance $d(C)$ is completely determined by

$d(C)=\left\{
\begin{array}{ll}
1, & $if $ i=0, \\

l+2$,$ & $if $lp^{s-1}+1\leq i\leq (l+1)p^{s-1}$ where $0\leq l\leq p-2, \\

(t+1)p^{k}$,$ &
\begin{array}{l}
$if $p^{s}-p^{s-k}+(t-1)p^{s-k-1}+1\leq i\leq p^{s}-p^{s-k}+tp^{s-k-1}$,$\\
$where $1\leq t\leq p-1$, and $1\leq k\leq s-1,%
\end{array}
\\
0, & $if $i=p^{s}.

\end{array}%
\right. $
\vskip 4pt

It is easy to verify that the cyclic codes of length $p^s$ over ${\cal R}=\mathbb{F}_{p^{m}}+u\mathbb{F}_{p^{m}}$ are precisely the ideals of the residue ring ${\cal R}_1=\frac{{\cal R}[x]}{\left\langle x^{p^s}-1\right\rangle}$. The following lemma is easy to verify.

\vskip 6pt
\noindent{\bf Lemma 3.2.}{
For any positive integer $n$, $(x-1)^{p^n} = x^{p^n}-1 \in {\cal R}[x]$. Moreover, $x-1$ is nilpotent in ${\cal R}_{1}$ with nilpotency index $p^s$.}

\vskip 6pt
\noindent{\bf Lemma 3.3.}{
Let $f(x) \in {\cal R}_{1}$. Then $f(x)$ can be uniquely written as
$$f(x)=\sum_{i=0}^{p^s-1}a_i(x-1)^i=f_1(x)+uf_2(x)=a_{0,0}+\sum_{i=1}^{p^s-1}a_{0,i}(x-1)^i+u\sum_{i=0}^{p^s-1}a_{1,i}(x-1)^i,$$
where $a_i=a_{0,i}+ua_{1,i}\in{\cal R} $, $a_{0,i},a_{1,i} \in \mathbb{F}_{p^m}$ and $f_1(x),f_2(x) \in \mathbb{F}_{p^m}[x]$. Furthermore, $f(x)$ is invertible if and only if $a_{0,0}=0$.
}

\vskip 6pt
{\it Proof.}{ The representation of $f(x)$ follows from the fact that it can be viewed as a polynomial of degree less than $p^s$ over ${\cal R}[x]$. Each coefficient $a_i$ of $f(x)$ is an element of ${\cal R}$, that can be expressed uniquely by $a_{0,i},a_{1,i} \in \mathbb{F}_{p^m} $ as $a_i = a_{0,i} + ua_{1,i}$. Expressing $f(x)$ in this representation, the last assertion follows from the fact that $u$ and $x-1$ are both nilpotent in ${\cal R}_{1}$. $\square$ }

\vskip 6pt
\noindent{\bf Theorem 3.4 \cite{dinh2010}}{
The ring ${\cal R}_{1}$ is a local ring with the maximal ideal $\left\langle u,x-1\right\rangle $, but it is not a chain ring. Cyclic codes of length $p^s$ over $\mathbb{F}_{p^{m}}+u\mathbb{F}_{p^{m}}$,
i.e., ideals of the ring ${\cal R}_{1}$, are

\begin{itemize}
\item Type $1$ (trivial ideals): $\left\langle 0\right\rangle $ , $\left\langle1\right\rangle $.

\item Type $2$ (principal ideals with nonmonic polynomial generators): $\left\langle u(x-1)^{i}\right\rangle$, where $0\leq i\leq p^{s}-1$.

\item Type $3$ (principal ideals with monic polynomial generators): $\left\langle(x-1)^{i}+u(x-1)^{t}h(x)\right\rangle $, where $1\leq i\leq p^{s}$ $-1$, $0\leq t<i$, and either $h(x)$ is $0$ or $h(x)$ is a unit.

\item Type $4$ (nonprincipal ideals):  $\left\langle
(x-1)^{i}+u(x-1)^{t}h(x),u(x-1)^{\omega}\right\rangle $, where $1\leq
i\leq p^{s}-1$, with $h(x)$ as in Type $3$, and $deg(h(x))\leq \omega -t-1$.
\end{itemize}
}

\section{Hamming Distance of Cyclic Codes}

As we mentioned in Section 3 the cyclic codes over ${\cal R}$ are precisely the ideal of the ring ${\cal R}_1$. In order to compute the Hamming distances of all cyclic codes over ${\cal R}$, we count the Hamming distance of the ideals of the ring ${\cal R}_1$ (Type 2, 3, 4).

\vskip 6pt
\noindent{\bf Proposition 4.1}{
Let $C =\left\langle u(x-1)^i\right\rangle \subsetneq {\cal R}_{1} $ be a cyclic code of length $p^s$ over ${\cal R}$ for some $0\leq i \leq p^s-1$. Then its minimum Hamming distance is completely determined by
\begin{equation}
d(C)=\left\{
\begin{array}{ll}
1, & $if $i=0, \\

l+2$,$ & $if $lp^{s-1}+1\leq i\leq (l+1)p^{s-1}$, where $0\leq l \leq p-2$,$ \\

(t+1)p^{k}, &
\begin{array}{l}
$if $p^{s}-p^{s-k}+(t-1)p^{s-k-1}+1\leq i\leq p^{s}-p^{s-k}+tp^{s-k-1}$,$\\

$where $1\leq t\leq p-1$, and $1\leq k\leq s-1$.$
\end{array}

\end{array}
\right.
\end{equation}
}
\vskip 6pt
{\it Proof.}{ For $i=0$ then $C=\left\langle u \right\rangle$, it follows that $d(C)=1$.\\
For $i \in \left\lbrace 1, 2,..., p-1\right\rbrace$, we have $\frac{\mathbb{F}_{p^{m}}+u\mathbb{F}_{p^{m}}}{\left\langle
u\right\rangle }\simeq \mathbb{F}_{p^{m}}$ then the ideals $\left\langle
u(x-1)^{i}\right\rangle $ of ${\cal R}_1$, are exactly the set of the ideals $\left\langle (x-1)^{i}\right\rangle $ of $\frac{\mathbb{F}_{p^{m}}[x]}{\left\langle x^{p^{s}}-1\right\rangle }$ multiplied by $u$. Hence, the Hamming distance of $\left\langle u(x-1)^{i}\right\rangle $ follows from Theorem~3.1. $\square$
}

In order to compute the minimum Hamming distances of those codes for the rest cases (Type 3 and 4),  we need the following very useful lemma.

\vskip 6pt
\noindent{\bf Lemma 4.2}{
\bigskip Let $c(x)=a(x)+ub(x)$ be an nonzero polynomial in ${\cal R}_{1}$, where $a(x),b(x)\in \mathbb{F}_{p^{m}}[x]$. Then we have
$$wt(c(x))\geq \max \{wt(a(x)),wt(b(x))\}.$$

\vskip 6pt
{\it Proof.}{
By Lemma 3.3, $c(x)$ can be written as
\begin{align*}
c(x) &=\sum_{i=0}^{p^s-1}(a_i+ub_i)x^i=a(x)+ub(x)=\sum_{i=0}^{p^s-1}a_ix^i+u\sum_{i=0}^{p^s-1}b_ix^i.
\end{align*}
Now suppose $wt(a(x))=d$, then there exist exactly $d$ nonzero $a_{i_{k}}\in \mathbb{F}_{p^{m}}$, where ${k}\in \{1,2,...,d\}$. Note that $c_{i}=a_{i}+ub_{i}$ and $c_i=0$ if and only if $a_i=b_i=0$. Hence $c_{i_k}\neq 0$ \ $\forall {i}\in \{1,2,...,d\}$. Therefore, $wt(c(x))\geq d=wt(a(x))$. We can show that $wt(c(x))\geq wt(b(x))$ by using the similar way.

Since $wt(c(x)) \geq wt(a(x))$ and $wt(c(x)) \geq wt(b(x))$, we get $wt(c(x))\geq \max \{wt(a(x),wt(b(x)\}$.  $\square$}

\vskip 6pt
\noindent{\bf Proposition 4.3}{
Let $C=\left\langle(x-1)^{i}+u(x-1)^{t}h(x)\right\rangle $ be a cyclic code of length $p^s$ over ${\cal R}$ for $1 \leq i \leq p-1$. Then the Hamming distance of $C$ is
\begin{equation}
d(C)=\left\{
\begin{array}{ll}
l+2$,$ & $if $lp^{s-1}+1\leq i\leq (l+1)p^{s-1}$, where $0\leq l\leq p-2$,$ \\

(t+1)p^{k}$,$ &
\begin{array}{l}
$if $p^{s}-p^{s-k}+(t-1)p^{s-k-1}+1\leq i\leq p^{s}-p^{s-k}+tp^{s-k-1}$,$\\ $where $1\leq t\leq p-1$, and $1\leq k\leq s-1$.$
\end{array}
\\

\end{array}%
\right.
\end{equation}

}

\vskip 6pt
{\it Proof.}{
Let $c(x)$ be an arbitrary nonzero element of $C$, then $c(x)$ can be written as $$
c(x)=(a(x)+ub(x))((x-1)^{i}+u(x-1)^{t}h(x)),$$ \ where $a(x),b(x)\in \mathbb{F}_{p^m}[x]$. We consider $c(x)$ in two cases.

Case 1:  $a(x)\neq 0$. Then we have
\begin{align*}
c(x) &=(a(x)+ub(x))((x-1)^{i}+u(x-1)^{t}h(x))\\
     &=a(x)(x-1)^{i}+u(b(x)(x-1)^{i}+a(x)(x-1)^{t}h(x))\\
     &=a(x)(x-1)^{i}+uh^\prime(x)
\end{align*}
where $h(x),h^\prime(x)\in \frac{\mathbb{F}_{p^{m}}[x]}{\left\langle x^{p^{s}}-1\right\rangle }$. By Lemma $4.2$, we obtain that
\begin{align*}
wt(c(x)) & \geq \max  \lbrace wt(a(x)(x-1)^{i}),wt(h^ \prime(x))\rbrace \geq wt(a(x)(x-1)^{i}).
\end{align*}

Case 2: $a(x)=0$. In this case, $b(x)\ne 0$. We have $c(x)=ub(x)(x-1)^{i}$, and it is easy to see that $wt(c(x))=wt(b(x)(x-1)^i)$.

We have shown that for every nonzero element $c(x)$ in $C$, there exist an element $d(x)$ in the ideal  $\left\langle (x-1)^{i}\right\rangle$ of the ring $\frac{\mathbb{F}_{p^m}[x]}{\left\langle x^{p^s}-1\right\rangle}$, such that $wt(c(x)) \geq wt(d(x))$. Therefore, $d(C) \geq d(\left\langle (x-1)^{i}\right\rangle)$.

On the other hand we have that $$u(x-1)^{i}=u((x-1)^{i}+u(x-1)^{t}h(x)) \in C$$
then $ \left\langle u(x-1)^i \right\rangle \subseteq C$. Hence, $d(\left\langle (x-1)^i \right\rangle)=d(\left\langle u(x-1)^i \right\rangle)\ge d(C)$, and we get $d(C)=d(\left\langle (x-1)^{i}\right\rangle)$. The rest of the proof follows from Theorem $3.1$ and the discussion above.$\square$
}

\vskip 6pt
\noindent{\bf Proposition 4.4}{
Let $i$ and $\omega$ be two  positive integers such as described in Theorem 3.4 and let $$C=\left\langle
(x-1)^{i}+u(x-1)^{t}h(x),u(x-1)^{\omega}\right\rangle. $$
Then the Hamming distance of $C$ is given by
\begin{equation}
d(C)=\left\{
\begin{array}{ll}

l+2$,$ & $if $lp^{s-1}+1\leq \omega \leq (l+1)p^{s-1}$, where $0 \leq l\leq p-2$,$ \\

(t+1)p^{k}, &
\begin{array}{l}
$if $p^{s}-p^{s-k}+(t-1)p^{s-k-1}+1\leq \omega \leq p^{s}-p^{s-k}+tp^{s-k-1}$,$\\

$where $1\leq t\leq p-1$, and $1\leq k\leq s-1$.$
\end{array}
\\

\end{array}%
\right.
\end{equation}
}

\vskip 6pt
{\it Proof.}{
We can easily see that $\omega <i$, otherwise $C$ \ is a principal ideal.

Let $c(x)$ be an arbitrary nonzero element of $C$, then $c(x)$ can be represented as $$%
c(x)=(a(x)+ub(x))((x-1)^{i}+u(x-1)^{t}h(x))+u(x-1)^{\omega }g(x)$$
where $a(x),b(x)$, and $g(x)$ are elements of  $\frac{\mathbb{F}_{p^{m}}[x]}{\left\langle x^{p^{s}}-1\right\rangle }$. We consider two cases.

Case 1: $a(x) \neq 0$. Then we can write $c(x)$ as
\begin{align*}
c(x) &=(a(x)+ub(x))((x-1)^{i}+u(x-1)^{t}h(x))+u(x-1)^{\omega }g(x) \\
	 &=a(x)(x-1)^{i}+u(b(x)(x-1)^{i}+(x-1)^{\omega }g(x))\\
     &=a(x)(x-1)^{i}+uh^\prime(x)
\end{align*}
 where $h^\prime(x)\in \frac{\mathbb{F}_{p^{m}}[x]}{\left\langle x^{p^{s}}-1\right\rangle }$, using the Lemma $4.2$ it follows that $wt(c(x))\geq wt(a(x)(x-1)^{i}) \geq d(\left\langle (x-1)^{i}\right\rangle)$.\\
Because $\left\langle (x-1)^{i}\right\rangle \subsetneq \left\langle (x-1)^{\omega}\right\rangle$ we have $wt(c(x)) \geq d(\left\langle
(x-1)^{\omega}\right\rangle)$.

\vskip 4pt
Case 2: $a(x)=0$ then we can written $c(x)$ as
\begin{align*}
c(x) &=u(b(x)(x-1)^{i}+(x-1)^\omega g(x))\\
	 &=u(x-1)^\omega (b(x)(x-1)^{i-\omega}+g(x))\\
	 &=u(x-1)^\omega h^\prime(x),
\end{align*}
where $h^\prime(x)$ is nonzero polynomial in $\frac{\mathbb{F}_{p^m}[x]}{
\left\langle x^{p^{s}}-1\right\rangle }$, and therefore we have
$$
wt(c(x))\geq wt((x-1)^{\omega }h'(x)).
$$
We have shown that for any nonzero element in $C$, there exist an element $e(x)$ in the ideal $\left\langle (x-1)^{\omega } \right\rangle$ such that $wt(c(x)) \geq wt(e(x))$. Therefore, $d(C) \geq d(\left\langle (x-1)^{\omega } \right\rangle)$.

On the other hand we have $\left\langle u(x-1)^{\omega }\right\rangle \subsetneq C$, Hence, the proof follows from Theorem~3.1 and the discussion above.$\square$
}

Now we can summarize the above results as the following theorem.

{\bf Theorem 4.5}{
The minimum Hamming distances of all cyclic codes of length $p^s$ over ${\cal R}$ are given as follows:
\begin{itemize}
\item For codes of $\left\langle0\right\rangle$, $\left\langle1\right\rangle$, their minimum Hamming distance are 0, 1 respectively.
\item For codes of $\left\langle u(x-1)^i\right\rangle$ where $0 \leq i\leq p^s-1$, their minimum Hamming distance are given in (1).
\item For codes of $\left\langle(x-1)^{i}+u(x-1)^{t}h(x)\right\rangle$, where $1\leq i\leq p^{s}$ $-1$, $0\leq t<i$, and either $h(x)$ is $0$ or $h(x)$ is a unit, their minimum Hamming distance are given in (2).
\item For codes of $\left\langle
(x-1)^{i}+u(x-1)^{t}h(x),u(x-1)^{\omega}\right\rangle $, where $1\leq
i\leq p^{s}-1$, with $h(x)$ as in Type $3$, and $deg(h)\leq \omega -t-1$, their minimum Hamming distance are given in (3).
\end{itemize}

\section{Hamming Distance of Constacyclic Codes}
In the last of the paper, we focus on the minimum Hamming distance of constacyclic codes over ${\cal R}$.

Let $\alpha$ be nonzero elements of the field $\mathbb{F}_{p^m}$, and $\beta$ be an  elements of the field $\mathbb{F}_{p^m}$, then $\lambda=\alpha+u\beta$ is a unit of ${\cal R}$. Therefore, there are $p^m(p^m-1)$ constacyclic codes corresponding to the units $\lambda$, for $\beta \neq 0$ the Hamming distances of $(\alpha+u\beta)$-constacyclic codes were studied in \cite{dinh2010}. In the reminder of the section, we address the $\alpha$-constacyclic codes by constructing one-to-one correspondence between cyclic and $\alpha$-constacyclic codes to apply our results to $\alpha$-constacyclic codes.

Note that $\alpha$ is a nonzero element of the finite field $\mathbb{F}_{p^m}$, then we have $\alpha^{p^m}=\alpha$. By the Division Algorithm, there exist nonnegative integers $\alpha_q$, $\alpha_r$ such that $s=\alpha_qm+\alpha_r$ with $0 \leq \alpha_r \leq m-1$. Let $\alpha_0 = \alpha^{-p^{(\alpha_q+1)m-s}}=\alpha^{-p^{m-\alpha_r}}$. Then $\alpha_0^{p^s}=\alpha^{-p^{(\alpha_q+1)m}}=\alpha^{-1}$. This observation leads to the following proposition.

\vskip 6pt
\noindent
{\bf Proposition 5.1}{
Let $\varphi$ be the map $\varphi : \frac{{\cal R}[x]}{\left\langle x^{p^s}-1 \right\rangle}  \longrightarrow  \frac{{\cal R}[x]}{\left\langle x^{p^s}-\alpha \right\rangle}$, given by $\varphi(f(x))=f(\alpha_0x)$, Then $\varphi$ is a ring isomorphism, and it is Hamming weight preserving.
}

\vskip 6pt
{\it Proof.}{
Let $f(x),g(x) \in {\cal R}[x]$, then
$$f(x)\equiv g(x)\pmod{x^{p^s}-1} \Leftrightarrow  f(x)=g(x)+h(x)(x^{p^s}-1),$$
where  $h(x) \in {\cal R}[x]$, which is equivalent to
$$f(\alpha_0x)=g(\alpha_0x)+h(\alpha_0x)(\alpha_0^{p^s}x^{p^s}-1)=g(\alpha_0x)+\alpha^{-1}h(\alpha_0x)(x^{p^s}-\alpha),$$
which is equivalent also to $f(\alpha_0x)\equiv g(\alpha_0x)\pmod {x^{p^s}-\alpha}$. This means that for $f(x),g(x)\in \frac{{\cal R}[x]}{\left\langle x^{p^s}-1\right\rangle}$, $\varphi(f(x))=\varphi(g(x))$ if and only if $f(x)=g(x)$. This implies that $\varphi$ is well-defined. It is easy to verify that  $\varphi$ is an ring isomorphism and weight-preserving.$\square$
}

The above ring isomorphism $\varphi$ provides a one-to-one correspondence between cyclic and $\alpha$-constacyclic code of length $p^s$ over ${\cal R}$. Hence we have the following theorem.

\vskip 6pt\noindent
{\bf Theorem 5.2}{
Assume the notation give above. Then the minimum Hamming distances of all $\alpha$-constacyclic codes of length $p^s$ over ${\cal R}$ are given as follows.
\begin{itemize}
\item For the codes of $\left\langle0\right\rangle$, $\left\langle1\right\rangle$, their minimum Hamming distances are $0, 1$ respectively.
\item For the codes of $\left\langle u(\alpha_0x-1)^i\right\rangle$ where $0 \leq i\leq p^s-1$, their minimum Hamming distances are given in (1).
\item For the codes of $\left\langle(\alpha_0x-1)^{i}+u(\alpha_0x-1)^{t}h(x)\right\rangle$, where $1\leq i\leq p^{s}$ $-1$, $0\leq t<i$, and either $h(x)$ is $0$ or $h(x)$ is a unit, their minimum Hamming distances are given in (2).
\item For the codes of $\left\langle
(\alpha_0x-1)^{i}+u(\alpha_0x-1)^{t}h(x),u(\alpha_0x-1)^{\omega}\right\rangle $, where $1\leq
i\leq p^{s}-1$, with $h(x)$ as in Type $3$, and $deg(h)\leq \omega -t-1$, their minimum Hamming distances are given in (3).
\end{itemize}
}

\vskip 6pt
\noindent
{\bf Remark 5.3}{ As pointed out in Section 3, $\alpha$-constacyclic codes of length $p^s$ over ${\cal R}$ are ideals of the ring ${\cal R}_\alpha$, which are not chain rings. For the cases ${\cal R}_{\alpha+u\beta}$ where $\beta \not =0$. In \cite{dinh2009} Dinh determined the Hamming distance for ${\cal R}_{\alpha+u\beta}$ where $p=2$. In \cite{dinh2010} Dinh extended these results for all prime numbers. The aim of this paper is the determination of the Hamming distance for the rest cases ($\beta=0$), where ${\cal R}_\alpha$ is a local ring, but not a chain ring.

When $p=2$, Theorem 5.2 and Theorem 5.1 of \cite{dinh2009}  complement each other in the sense that they deal with $\lambda$-constacyclic over ${\cal R}$. Furthermore, for any prime number Theorem 5.2 and Theorem 4.4 of \cite{dinh2010} complement each other in the sense that they deal with $\lambda$-constacyclic over ${\cal R}$.
}

\end{document}